# Selective Noise Resistant Gate

Jonatan Zimmermann[1], Paz London[1], Yaniv Yirmiyahu[1],
Fedor Jelezko[2], Aharon Blank[3], David Gershoni[1]

[1]*Physics Department and Solid State Institute, Technion - Israel Institute of Technology, Israel*

[2]*Institut für Quantenoptik, Universität Ulm, Germany*

[3]*Schulich Faculty of Chemistry, Technion - Israel Institute of Technology, Israel*

Realizing individual control on single qubits in a spin-based quantum register is an ever-increasing challenge due to the close proximity of the qubits resonance frequencies. Current schemes typically suffer from an inherent trade-off between fidelity and qubits selectivity. Here, we report on a new scheme which combines noise protection by dynamical decoupling and magnetic gradient based selectivity, to enhance both the fidelity and the selectivity. With a single nitrogen-vacancy center in diamond, we experimentally demonstrate quantum gates with fidelity $= 0.9 \pm 0.02$ and a $50 \, kHz$ spectral bandwidth, which is almost an order of magnitude narrower than the unprotected bandwidth. Our scheme will enable selective control of an individual nitrogen-vacancy qubit in an interacting qubits array using relatively moderate gradients of about $1 \, mG/nm$.

## I. Introduction

Quantum information processing is widely believed to outperform classical analogs [1] and is expected to overwhelmingly impact science and technology. It promises exponential speedup in certain computational tasks such as quantum chemistry simulations [2], solving linear equations [3], machine learning [4], and prime factoring [5]. To realize these promises, a quantum processor requires a network of qubits, which can be independently initialized, controlled, and readout. The network nature of the processor is achieved via strong coherent coupling between the qubits in the array.

One of the most promising qubit types in solid-state architectures is electronic spins [6]. To ensure a robust two qubit gate, the dipolar interaction must be significantly stronger than the decoherence process. At room temperature environment, this condition sets a limit to the maximal separation between spins of ~10 nm. This separation is much smaller than the radiation diffraction limit (optical or RF), thus limiting the ability to focus the driving field on a single site, thereby hindering individual control of each qubit. The common solution to the desired distinguishability is based on spectral means. I.e., having a different energy splitting for each qubit, thus allowing for selective manipulation by tuning the driving field frequency to the resonant frequency of a specific qubit while being far detuned from all the other ones.

Spectral discrimination can be achieved either by an inherent difference between the qubits (e.g. their orientation in the hosting lattice [7]) or more generally, by an external physical field gradient [8-13]. The gradient-induced frequency separation needs to be larger than the qubits' natural linewidth. In many experimental scenarios, this proves to be a challenging task. When the spectral separation between qubits is comparable to or smaller than the linewidth, we are facing an inherent tradeoff between selectivity and fidelity. On one hand, a selective narrowband operation suffers from unwanted qubit resonance jitter and drifts, which reduces its fidelity. On the other hand, a broadband operation is insensitive to frequency noise but parasitically drives the other qubits, thus destroying the selectivity.

In this work, we develop and demonstrate a protocol that maintains both selectivity and high fidelity. We combine noise canceling dynamical decoupling (DD) sequences [14-17] and pulsed magnetic field gradient [8] into the driving scheme. By that, the spins coherence times are prolonged while the spectral based selectivity is retained, despite the deleterious nature of the DD schemes. Our scheme does not require any additional control resources.

We proceed by describing the theory of our scheme applied to a general spin 1/2 system. We then verify the scheme experimentally using the electronic spin of a single nitrogen-vacancy (NV) center in diamond [18]. NV center stands as a promising physical system for quantum information processing, sensing, and communication, due to its outstanding qubit properties: long coherence time, optical initialization and readout, high spatial qubits density, and ease of quantum control [19].

## II. Protection scheme

Our model system is a single spin qubit, whose quantization axis and eigenstates lay on the z-axis. The qubit is subjected to external driving in the XY plane. In the rotating frame, the Hamiltonian of a unitary rotation around an axis with azimuthal phase angle $\phi$ can be described as:

$$H = B_Z \cdot S_Z + \Omega \cdot (\cos\phi\, S_X + \sin\phi\, S_Y), \quad (1)$$

where $B_Z$ is offset field, $\Omega$ is the Rabi frequency, and $\mathbf{S} = (S_X, S_Y, S_Z)$ is the spin vector operator. For the simple case when the detuning is negligible compared to the driving, we denote the operator $\theta_{(\cos\phi\, X + \sin\phi\, Y)} = e^{-i\theta(\cos\phi\, S_X + \sin\phi\, S_Y)}$ as describing the evolution of the spin due to pulse with time duration $\tau$, such that $\theta = \tau \cdot \Omega$.

Generally, the driving frequency may be detuned from the resonant frequency of the qubit by $\Delta_Z$ and additional noise ($\delta$) may interact with the spin. Under the secular approximation, both the detuning and the noise contribute an additional Hamiltonian term $H_Z \propto S_Z$. For solid-state defects, the dominant noise comes from a slowly fluctuating magnetic field from the nearby nuclear spins [20], therefore one can assume it is time independent during gate execution. Using the above model and assumptions, a realistic gate is represented by

$$\theta_{(\cos\phi\, X + \sin\phi\, Y),\ \Delta_Z + \delta} = e^{-i\tau(\Omega(\cos\phi\, S_X + \sin\phi\, S_Y) + (\Delta_Z + \delta)S_Z)}. \quad (2)$$

We choose to focus on a $\pi_X$ gate (i.e. $\theta = \pi$ and $\phi = 0$). Useful implementation for a single qubit gate has both high fidelity and low spectral bandwidth, i.e. high selectivity. The fidelity ($F$) is defined as the output state fidelity after applying a resonant $\pi_X$ gate on the $|m_s = 0\rangle$ state and the bandwidth ($BW$) is defined as the spectral detuning required to achieve state fidelity of less than 0.1 (corresponds to $\Delta_Z = \Omega$ for ideal Rabi driving). See Appendix B.I for the definitions and derivation.

When the gradient-induced separation is much larger than the natural linewidth $\Delta_Z \gg 1/\pi T_2^*$, it is convenient to choose Rabi Frequency $\Omega$ such that $1/\pi T_2^* \ll \Omega \ll \Delta_Z$, thus fulfilling the two objectives of selectivity and high fidelity. Here, we expand the viable gradient intensities to the regime where $\Omega < \Delta_Z \leq 1/\pi T_2^*$, by extending the time scale to $T_2$, the protected decoherence time, such that $1/\pi T_2 < \Omega < \Delta_Z$. The development of the selective noise resistance gate (SNRG) scheme is described in Fig 1, in three consecutive steps. In each step, we present a scheme by displaying the sequence of its operators and the time evolution of an on- and off-resonance spins on their respective Bloch sphere representation.

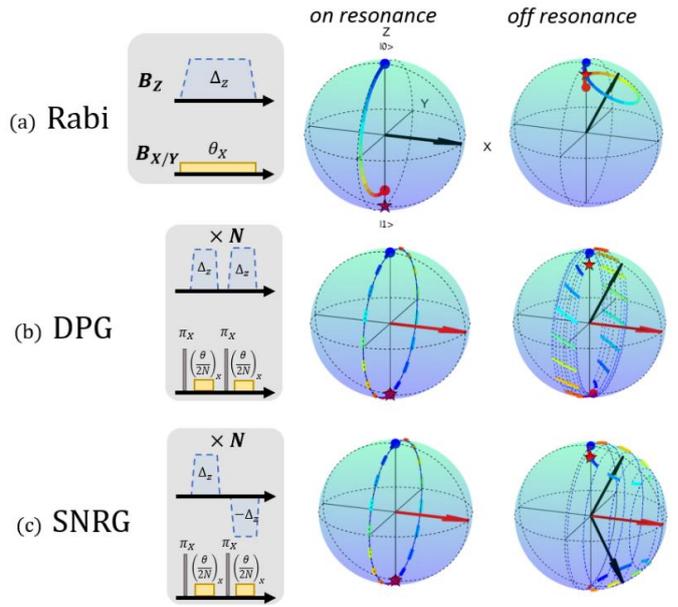

*Fig 1: Various schemes for single qubit control under noisy environment.* **Left column:** *schematic description of the temporal evolution of the ($B_Z$) and the microwave driving ($B_{X/Y}$).* **Middle column:** *The resulting spin dynamics for resonant control, depicted on its Bloch sphere.* **Right column:** *The resulting spin dynamics for detuned control with $\Delta_Z = 2\Omega$, depicted on its Bloch sphere.*

*(a) Simple Rabi gate, the resonantly controlled spin rotates, and the detuned spin remains relatively intact. Both spins experience decoherence. (b) Dynamically protected gate (DPG), both spins experience rotation with much slower decoherence. Thus, selectivity is lost. (c) The SNRG scheme. The resonant spin rotates with slower dephasing, while the detuned spin does not.*

*The red stars indicate the final state of for an ideal gate (without noise). There are two types of contributions to the spin dynamics: Rabi driving segments (represented by colored arcs) and dynamical decoupling driving by $\pi$ pulses (depicted by thin dashed arcs).*

Scheme (a) is a simple, continuous driving Rabi gate with $\Omega < 1/\pi T_2^*$. The on resonance ($\Delta_Z = 0$) spin performs a rotation around $\hat{X}$ axis (Fig 1a middle), and the off-resonance spin ($\Delta_Z = 2\Omega$) experiences a rotation around the tilted axis: $\Omega\hat{X} + \Delta_Z\hat{Z}$ (Fig 1a right). Both spins experience decoherence, characterized by $T_2^*$. The effective bandwidth in this scheme is approximately $BW \approx \Omega$, and the fidelity of the resonantly controlled spin is given by $T_2^*$ as well (see [21]). Thus, in this scheme better selectivity leads to lower fidelity, as both $F$ and $BW$ decrease with $\Omega$.

Scheme (b) describes a dynamically protected gate (DPG) [14-17]. Here, dynamically decoupling (DD) pulses are applied in parallel to the gate operation, in order to reduce the effect of the noisy environment. The DD pulses narrow the effective linewidth from $1/\pi T_2^*$ towards the protected decoherence rate $1/\pi T_2$. This modification is done by dividing the control gate to $N$ multiple segments, while interleaving a standard DD sequence at the waiting times between the segments (Fig 1b left). Each DD pulse is a short period $\pi$ pulse, with a pulse rate $\Omega_{DD}$ far exceeding the decoherence rate ($1/\pi T_2^*$). The magnetic gradient is set to be constant throughout the gate. The DD cancels out the effects of the noise $\delta$, but it also eliminates the detuning $\Delta_Z$. Thereby, both on and off resonance spins rotate (see Fig 1b middle and right). It is possible to show, that the gate dynamics converge to driving with a diminishing $S_Z$ term when $N \to \infty$. For example, the dynamics of the Carr – Purcelll – Meiboum - Gill (CPMG) DD sequence is given by eq. 3 as explained in Appendix B.II:

$$\left(\pi_X \cdot \left(\frac{\theta}{2N}\right)_{X,\ \Delta_Z} \cdot \pi_X \cdot \left(\frac{\theta}{2N}\right)_{X,\ \Delta_Z}\right)^N \xrightarrow{N \to \infty} \theta_{X,\ 0}. \quad (3)$$

Thus, this scheme clearly improves the fidelity but suffers from low selectivity under magnetic field gradient, since the Zeeman interaction is refocused.

Scheme (c) describes the Selective Noise Resistant Gate (SNRG). This gate is based on the possibility to engineer the magnetic gradient, in such a way that it is not canceled by the DD pulses. The essential modification that we propose is to alternate the magnetic gradient between positive and negative values with each DD inversion pulse (Fig 1c left). This way, the $S_Z$ detuning term is conserved, as described for example for a CPMG sequence:

$$\left(\pi_X \cdot \left(\frac{\theta}{2N}\right)_{X,\ \Delta_Z} \cdot \pi_X \cdot \left(\frac{\theta}{2N}\right)_{X,\ -\Delta_Z}\right)^N = \theta_{X,\ \Delta_Z}. \quad (4)$$

Here, a resonant spin will rotate in a protected way due to magnetic noise cancellation (Fig 1c middle), while a detuned spin will rotate according to the generalized Rabi model with mitigated noise (Fig 1c right).

### III. Experimental setup

For the experimental demonstration, we use the electronic spin of an NV center in diamond. The qubit consists of a subsystem of the entire energy structure, $|m_s = 0\rangle$ and $|m_s = -1\rangle$ states with a splitting of about $1.8\ GHz$ between them. The microwave driving and the magnetic field gradient are applied by two crossing microwires, which are stretched on the diamond surface near the studied NV-center, as seen in Fig 2. Further details on the experimental apparatus can be found in Appendix A.

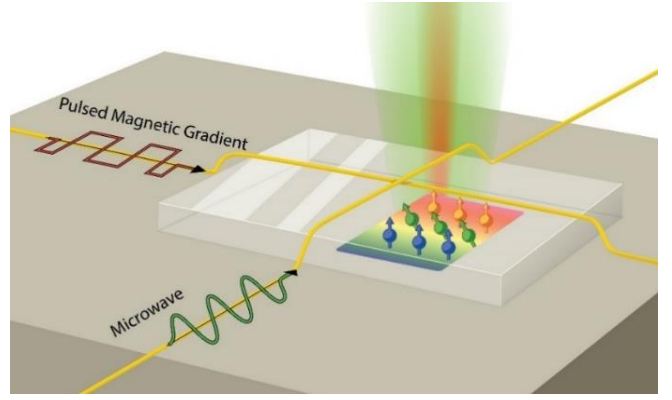

*Fig 2: Schematic of the experimental setup. The longitudinal magnetic field $B_Z$ is applied by an external permanent magnet (not shown) and a proximate microwire, while a second microwire is dedicated to the microwave driving.*

In any physical implementation of logical gates, the control pulses suffer from experimental imperfections, which rapidly accumulate and may degrade the system coherence. To reduce this problem we use robust pulse sequences based on multi-axis driving, such as XY-8 [22]. Multi-axis driving, give rise, however, to problems due to possible noncommutability between the desired control gate $\theta_{X,\Delta_Z}$, and off-axis DD pulses, such as $\pi_Y$ for example [14-17]. One way to circumvent this problem is to invert the microwave driving phase after each uncommutable DD operator (phase cycling). The operator equality of eq. 5 is explained in Appendix B.III:

$$\pi_Y \cdot (-\theta)_{X,\ -\Delta_Z} \cdot \pi_Y \cdot \theta_{X,\ \Delta_Z} = 2\theta_{X,\ \Delta_Z}. \quad (5)$$

Fig 3 illustrates the full SNRG scheme. Three parallel control channels comprise the scheme: spin initialization and readout using 532 nm laser, alternating longitudinal magnetic field $B_Z = B_0 + B_1 \cdot U(t)$ where $U(t)$ is a rectangular pulse train function (further details in Appendix B.IV) and microwave driving along $X$ and $Y$ axes, in the rotating frame.

Since the $B_Z$ field is switched during the sequence, the phase of the spin is accumulated with a changing rate:

$$\phi = \int_0^t B_Z(s)ds. \quad (6)$$

In order to continuously drive the spin, the microwave driving needs to follow the instantaneous axis (namely the phase) of the spin. Accordingly, such driving field is described by the frequency modulation (FM) equation:

$$B_X = \gamma^{-1}\Omega\ cos\left(2\pi \cdot \gamma \int_0^t B_Z(s)ds\right). \quad (7)$$

The derivation is brought in Appendix B.V.

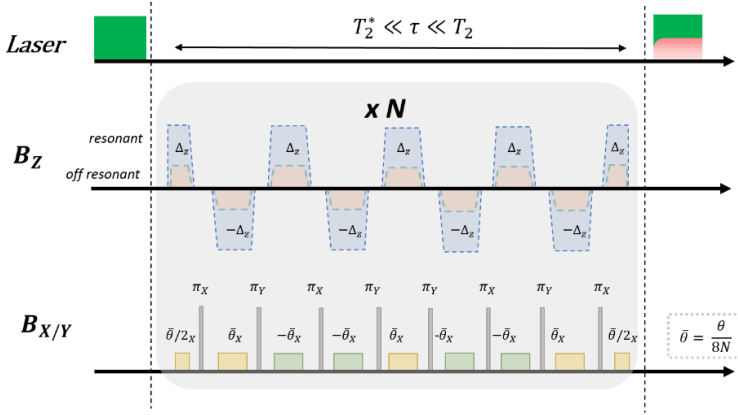

*Fig 3: SNRG sequence consisting of laser initialization (green), read-out pulse (pale red), a pulsed magnetic field gradient, and a modified XY-8 sequence. The $B_Z$ field as experienced by resonant and detuned spins are illustrated in pale blue and pale orange, respectively. The microwave frequency and phase of the driving field $B_x$ are controlled by frequency modulation and phase cycling (see text) to follow the instantaneous changes in the magnetic field for the resonant spin. Segments that correspond to driving along the X axis depicted in orange and the inverted driving in pale green.*

## IV. Results

We compare the fidelity and selectivity performance of the SNRG sequence with the simpler schemes (Rabi and DPG) shown in Fig 4. The comparison is performed by scanning the magnetic field gradient while fixing the microwave frequency dynamics, thus emulating various on/ off spin resonance possibilities.

Here, the Rabi frequency is $\Omega = 54\ kHz$ for the on resonance spin. In the continuous Rabi driving measurements (Fig 4. Left), low fidelity oscillations are measured - $F = 0.27 \pm 0.1$.

The decoherence is associated to the $^{13}$C nuclear spin bath surrounding the NV center, and quantified by the dephasing time $T_2^* = 5 \pm 1\ \mu s$. Hence, we model this magnetic bath influence using Ornstein - Uhlenbeck process. This process treats many spin-bath environment as an effective magnetic field with a correlation function $\langle B_{bath}(t)B_{bath}(0)\rangle = b^2 exp(-t/\tau_C)$, where $b$ is the coupling of the bath to the spin and $\tau_C$ is the correlation time determined by intrabath dipolar interaction [21][22]. We fitted the measured data with $b = 42\ kHz$, and $\tau_C = 230\ \mu s$, which reproduce the entire spin decoherence dynamics for all the driving detunings.

The middle panel shows the DPG scheme's ability to significantly reduce the effect of the noise and unfortunately also of the gradient induced detuning. Thus, the fidelity increases to $F = 0.9 \pm 0.02$ at the expense of the selectivity. This is achieved by applying XY-8 DD pulse sequence in parallel to the continuous Rabi driving. The experiment comprises an increasing number of XY-8 pulse cycles, where for each time step of $1\ \mu s$ an additional cycle is added. Each cycle consists of 8 $\pi$ pulses, with a time duration of $20\ ns$ per pulse and a delay of $125\ ns$ between the pulses. It is important to note that in contrast to the SNRG scheme, in the DPG scheme we leave the gradient and the microwave frequency unchanged during the entire evolution.

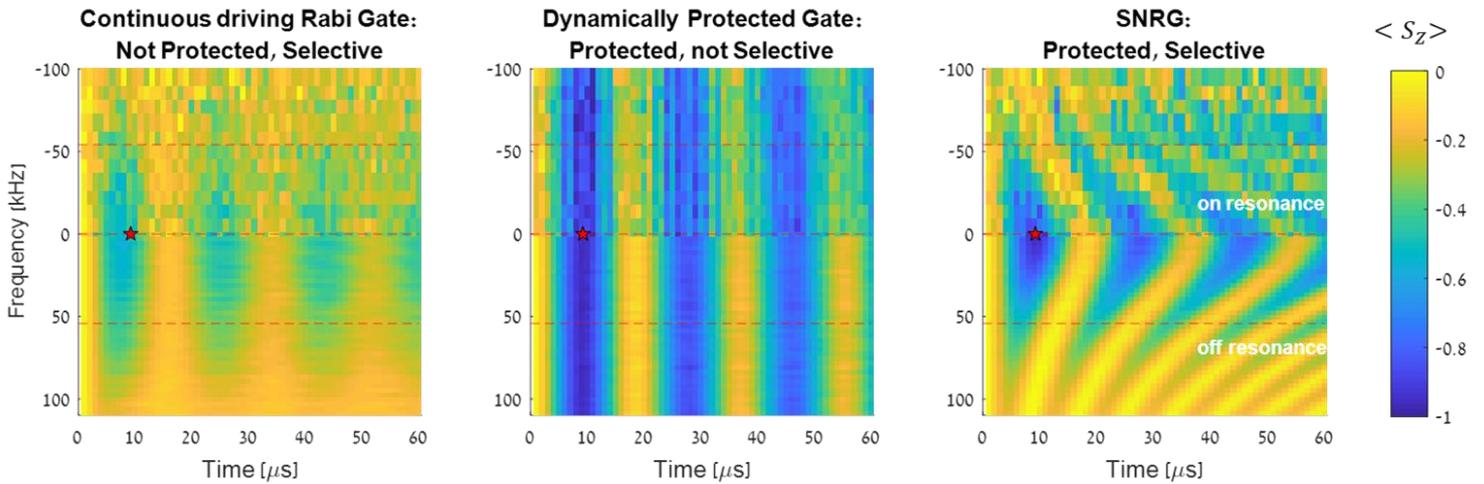

*Fig 4: Comparison between three driving schemes applied to a single NV center. The upper part (negative detuning) describes the measured data, while the lower part (positive detuning) describes model simulations. The comparison is presented by measuring $S_Z$ spin projection at the end of the scheme as a function of total pulse length and detuning between the driving frequency and the qubit resonance. The red stars indicate the location of a resonant $\pi$ pulse.*

The last panel shows the SNRG scheme performance, where DD is applied in conjunction with magnetic field modulation. The microwave frequency is given by a pulse train function, $f = f_0 + f_1 \cdot U(t) = 1800\ MHz + 180\ kHz \cdot U(t)$, which matches a specific magnetic field dynamics $B_Z = B_0 + B_1 \cdot U(t) = 380\ G + 70\ mG \cdot U(t)$. The relation between the driving frequency and the longitudinal magnetic field is determined by $f(t) = D - \gamma B_Z(t)$, where $D$ is the NV center zero-field splitting. To investigate the scheme performance, we scan $B_1$, the longitudinal field amplitude, between $30 - 100\ mG$. Therefore, the presented detuning in the vertical axis is the amplitudes difference, $f_1 - \gamma B_1$.

Using this scheme, we get significantly improved performance, much close to an ideal driving. A single $\pi$ pulse rotates $|m_s = 0\rangle$ to $|m_s = -1\rangle$ with state fidelity $F = 0.9 \pm 0.02$ and spectral bandwidth of $BW = 49 \pm 5\ kHz$. We attribute the residual infidelity of about 0.1 to the accumulation of experimental imperfections in our setup. In particular, the exact timing and duration of the short $\pi$ pulses, dominated by the jitter of our arbitrary waveform generator (AWG) device. In Fig 11 (Appendix C.II), a numerical simulation of the SNRG scheme is presented, which takes into account the imperfection in the DD pulses. We believe that better equipment with improved timing stability can push the fidelity closer to unity. The resulting SNRG bandwidth is limited only by the driving strength $\Omega$, and therefore can be further reduced.

In Fig 5, we extrapolate and predict how the SNRG will improve the driving performance for various driving Rabi frequencies. These calculations were carried out by Monte Carlo simulations, with the extracted noise parameters and DD pulses imperfections, as detailed in Appendix C.III.

We find that a fast degradation of the standard Rabi scheme fidelity occurs when the driving frequency decreases below the effective noise coupling $b$. At this extreme regime, the SNRG scheme improves the fidelity by more than 30-fold (Fig 5a, dashed arrow), and the selectivity is still limited by the driving Rabi frequency only - $BW = \Omega$ (Fig 5b). The limiting factors to SNRG performance are determined so far by the quality of the DD sequence.

## V. Discussion

Based on our measurements and simulations, we can assess the required magnetic gradient for a future NV based quantum register.

To date, we reached gate bandwidth of about $50\ kHz$, equivalent to $20\ mG$. Therefore, in order to reach a spectral separation of $\Delta_Z > 5\Omega$ (corresponding to $F < 10^{-3}$, as shown in Appendix B.I) between $10\ nm$ proximate spins (typical distance between coupled spins [24]), a magnetic gradient of $10\ mG/nm$ is currently required. Indeed, previous works demonstrated such separation ability [23][13], but with much higher gradient strengths which implies more complex fabrication procedure and experimental apparatus.

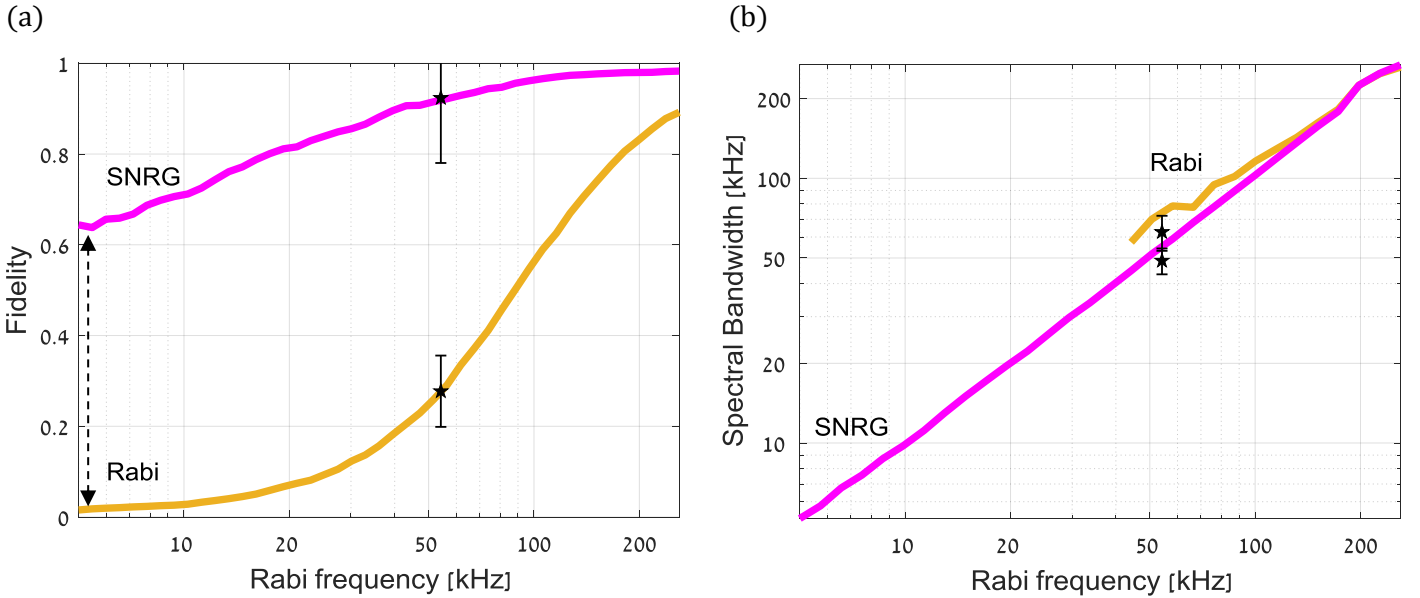

*Fig 5: Calculated SNRG enhancement of fidelity (a) and selectivity (b) over simple Rabi driving for different driving frequencies. The spectral bandwidth is displayed only for the cases when the on resonance fidelity is larger than 0.2. The black marks indicate the measured fidelity and selectivity for both schemes.*

We believe that with further, feasible improvements such as stabilization of the DD timing and duration, we will be able to improve our $BW$ by more than an order of magnitude reaching $BW = 5\ kHz$, equivalent to $2\ mG$. Thus, a moderate gradient of only $1\ mG/nm$ will be sufficient, expanding substantially the viable implementation alternatives.

The protocol presented here takes an important step towards solid-state spin-based quantum information processing. By combining two known practices: magnetic gradient induced spectral selectivity and dynamically protected gate, we managed to improve gate selectivity by more than an order of magnitude, essentially limited only by $T_1$. This improvement reduces the required magnetic gradient and opens the possibility for new experimental implementations. In particular, quantum register based on dipolar coupling could include parallel spins, thus expanding previous works [24] and allowing scaling up. Furthermore, the described scheme is readily extendable to 2D arrays and other atom-like systems such as trapped ions [25] and neutral atoms [26]. Another exciting application could be high-resolution imaging of dense spins system [27].

### Acknowledgments

The support of the Israeli Science Foundation (ISF) is gratefully acknowledged.

## Appendix A: Experimental Setup

The experiment is performed on a single NV center in a type IIa, electronic grade (nitrogen concentration < 5 ppb), single crystal diamond with a natural abundance of $^{13}$C (Element 6). The NV is addressed with green solid-state laser (gem 532, laser quantum), switched by an acousto-optical modulator (isomet), and focused on the diamond by oil-immersion microscope objective (Olympus). The photoluminescence is separated from the excitation light by a longpass dichroic beamsplitter (T565LPXR, chroma) and collected by a single photon counter module (SPCM-AQRH-14, Excilitas) through a confocal microscope setup. The photon detections are counted by a time tagging device (Time Tagger 20, Swabian Instruments). We scan the diamond by moving the objective with a nanopositioning stage (NPXYZ200Z25, NPOINT).

To isolate a single two-level system, we polarize the nuclear spin of the substitutional nitrogen atom to $m_I = -1$ using the excited state level-anti crossing (ESLAC), which occurs at a magnetic field of 510 G. In our experiment, the magnetic field is estimated to be 380 G and generated by an external, cylindrical, samarium-cobalt magnet (SmCo22LTC, Bunting). This field is sufficient for reasonable polarization due to the strong hyperfine interaction in the excited state [28].

The quantum gates are executed by microwave driving. The signals are generated through a single side band (SSB) IQ modulation with a 100 MHz intermediate frequency (IF) signal synthesized by 1.2 Gs/s arbitrary waveform generator (AWG 5014c, Tektronix) with a local oscillator from vector signal generator (E4438C, Agilent). The IF is fed into the IQ inputs of the signal generator. The AWG controls all the microwave operations during the driving schemes: switching, changing the amplitudes, frequencies, and phases of the driving, while the signal generator sets the carrier frequency (matching the qubit resonance frequency). After the mixing, the signal is sent through a 16W power amplifier (ZHL-16W-43+, minicircuits) to a 15 μm microwire placed in close proximity to the addressed NV center.

A key feature in the SNRG scheme is the alternating magnetic gradient during the quantum gate. This field is generated by current pulses obtained directly from the AWG and fed to a second 15 μm microwire on the diamond. The strength of the magnetic field is determined by measuring the resonance shift in ODMR experiments. We calibrate a linear relation of 0.93 MHz frequency shift per 1 V of AWG driving voltage.

Finally, we mention the software employed in our laboratory. We use Qudi [29], open source software, custom designed for performing quantum optics experiments. The software emphasizes modularity, flexibility, automatization, extended graphical interfaces, data recording, and other useful features.

# Appendix B: Additional theoretical details

## I. Fidelity and bandwidth measures for quantum gates implementation

In this work, various quantum gates implementations are designed to reach maximal fidelity and selectivity. Hence, we define two independent figures of merit to evaluate the performance of various schemes: output state fidelity and gate bandwidth. We denote a general single qubit operation as $U_{\Delta_Z}$ where $\Delta_Z$ is the detuning from resonance. The first measure, fidelity, is defined as the overlap between the ideal and realistic states after application of a resonant gate $F = <m_s = 1|\pi_X|m_s = 0>^2$. The bandwidth measure is defined as the required spectral detuning, $\Delta_Z$, to achieve output state fidelity of less than 0.1, $BW = max_{\Delta_Z}[F_{\Delta_Z} < 0.1]$, $F_{\Delta_Z} = <m_s = 1|\pi_{X,\Delta_Z}|m_s = 0>^2$.

For an ideal implementation of a $\pi$ pulse ($\pi_{X,\Delta_Z} = e^{-it_\pi(\Omega S_X + \Delta_Z S_Z)}$, where $t_\pi \cdot \Omega = \pi$) the fidelity is clearly equals to 1 when $\Delta_Z = 0$, and the bandwidth is approximately equal to the Rabi frequency $BW \approx \Omega$, as explained below.

The generalized Rabi theory for spin 1/2 driving gives us:

$$<1|\pi_{X,\Delta_Z}|0> = \frac{\Omega^2}{2(\Omega^2+\Delta_Z^2)} \cdot \left(1 - cos\left(\sqrt{\Omega^2+\Delta_Z^2} \cdot t_\pi\right)\right)$$
$$= \frac{\Omega^2}{2(\Omega^2+\Delta_Z^2)} \cdot \left(1 - cos\left(\pi\frac{\sqrt{\Omega^2+\Delta_Z^2}}{\Omega}\right)\right) \quad (B1)$$

where $t_\pi \cdot \Omega = \pi$.

In the NV center, we have a spin triplet system ($S = 1$) in the ground subsystem. In our study, we isolate an effective qubit between levels $m_S = 0$ and $m_S = -1$ and operate on it.

To conclude, the fidelity is:

$$F_{\Delta_Z} = \langle -1|\pi_{Z,\ \Delta_Z}|0\rangle^2 = \left(\frac{1}{2(1+r^2)} \cdot \left(1 - cos(\pi\sqrt{1+r^2})\right)\right)^2 \ ; \ r = \frac{\Delta_Z}{\Omega} \quad (B2)$$

as presented in Fig 6. The fidelity reaches to 0.1, at around $\Delta_Z \approx \Omega$, and at around $\Delta_Z \approx 5\Omega$ it reaches to $10^{-3}$.

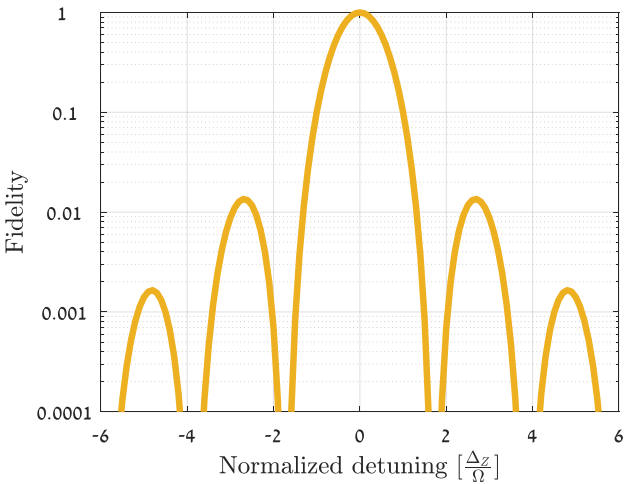

## II. Proof for CPMG based gate protection

To elucidate the principle behind Dynamically Protected Gate (DPG) [14-17], we use direct calculation. Assume a single qubit quantum gate, namely, a rotation of angle $\theta$ around $X$ axis. The simplest way to execute this gate, is a continuous driving for a time duration, $\tau$, with $\Omega_X = \theta/\tau$ Rabi frequency:

$$\theta_{X,\ 0} = e^{-i\tau\Omega_X S_X} \quad (B3)$$

The alternative scheme, DPG, proposes to eliminate the external coupling to $Z$ axis magnetic field - $\Delta_Z \cdot S_Z$. As described in the main text, the gate is divided into many segments and then interleaved with a standard dynamical decoupling (DD) scheme. For the CPMG scheme, there are $2N$ segments, each with a time duration of $\bar{\tau} = \frac{\tau}{2N}$, causing a rotation of $\bar{\theta} = \frac{\theta}{2N} = \bar{\tau} \cdot \Omega_X$ under zero $Z$ axis magnetic field. We will show that when $N \rightarrow \infty$, the DPG scheme executes protected quantum gate.

$$\left(\pi_X \cdot \bar{\theta}_{X,\ \Delta_Z} \cdot \pi_X \cdot \bar{\theta}_{X,\ \Delta_Z}\right)^N \stackrel{N\rightarrow\infty}{\Longrightarrow} \theta_{X,\ 0} \quad (B4)$$

We denote, the generalized rabi frequency as $\rho = \sqrt{\Omega_X^2 + \Delta_Z^2}$. A single cycle consists of 2 Rabi segments and 2 DD $\pi$ pulses. Under the described limit:

$$\pi_X \cdot (\bar{\tau} \cdot \Omega_X)_{X,\ \Delta_Z} \cdot \pi_X \cdot (\bar{\tau} \cdot \Omega_X)_{X,\ \Delta_Z} =$$

$$\begin{pmatrix} 0 & -i \\ -i & 0 \end{pmatrix} \cdot \begin{pmatrix} cos(\bar{\tau}\rho/2) & \frac{-i\,sin(\bar{\tau}\rho/2)\Omega_X}{\rho} \\ \frac{-i\,sin(\bar{\tau}\rho/2)\Omega_X}{\rho} & 0 \end{pmatrix} \cdot \begin{pmatrix} 0 & -i \\ -i & 0 \end{pmatrix} \cdot$$

$$\begin{pmatrix} cos(\bar{\tau}\rho/2) & \frac{-i\,sin(\bar{\tau}\rho/2)\Omega_X}{\rho} \\ \frac{-i\,sin(\bar{\tau}\rho/2)\Omega_X}{\rho} & 0 \end{pmatrix} =$$

$$\begin{pmatrix} -\frac{\Delta_Z^2 + cos(\bar{\tau}\rho)\Omega_X^2}{\rho^2} & \frac{(i\rho\,sin(\bar{\tau}\rho) + (cos(\bar{\tau}\rho)-1)\Delta_Z)\Omega_X}{\rho^2} \\ \frac{(i\rho\,sin(\bar{\tau}\rho) - (cos(\bar{\tau}\rho)-1)\Delta_Z)\Omega_X}{\rho^2} & -\frac{\Delta_Z^2 + cos(\bar{\tau}\rho)\Omega_X^2}{\rho^2} \end{pmatrix}$$

$$\stackrel{\bar{\tau}\rho \rightarrow 0}{\Longrightarrow} \begin{pmatrix} -\frac{\Delta_Z^2 + \left(1-\frac{(\bar{\tau}\rho)^2}{2}\right)\Omega_X^2}{\rho^2} & \frac{\left(i\rho(\bar{\tau}\rho) + \left(-\frac{(\bar{\tau}\rho)^2}{2}\right)\Delta_Z\right)\Omega_X}{\rho^2} \\ \frac{\left(i\rho(\bar{\tau}\rho) - \left(-\frac{(\bar{\tau}\rho)^2}{2}\right)\Delta_Z\right)\Omega_X}{\rho^2} & -\frac{\Delta_Z^2 + \left(1-\frac{(\bar{\tau}\rho)^2}{2}\right)\Omega_X^2}{\rho^2} \end{pmatrix}$$

$$= \begin{pmatrix} -\left(1-\frac{(\bar{\tau}\Omega_X)^2}{2}\right) & \left(i\bar{\tau} - \frac{\bar{\tau}^2\Delta_Z}{2}\right)\Omega_X \\ \left(i\bar{\tau} + \frac{\bar{\tau}^2\Delta_Z}{2}\right)\Omega_X & -\left(1-\frac{(\bar{\tau}\Omega_X)^2}{2}\right) \end{pmatrix}$$

$$= \begin{pmatrix} -cos(\bar{\tau}\Omega_X) & i\,sin(\bar{\tau}\Omega_X) \\ i\,sin(\bar{\tau}\Omega_X) & -cos(\bar{\tau}\Omega_X) \end{pmatrix}$$

$$= (2\bar{\tau} \cdot \Omega_X)_{X,\ 0} = 2\bar{\theta}_{X,\ 0} \quad (B5)$$

*Fig 6: Output state fidelity of single qubit gate as a function of the driving detuning.*

### III. Proof for XY based gate protection

A common problem with protected driving schemes is the noncommutability between the original driving and the added dynamical decoupling pulses. Thus, the DD pulses are not canceled out and may ruin the driving. Fortunately, this can be solved by altering the phases of the driving segments, so that they will accumulate constructively. The underlying example is a modification of $X$ axis driving with $Y$ axis DD. This scheme works, also when there is a slight $Z$ axis detuning, as brought below.

$$\pi_Y \cdot (-\theta)_{X,\ -\Delta_Z} \cdot \pi_Y \cdot \theta_{X,\ \Delta_Z} = 2\theta_{X,\ \Delta_Z} \quad (B6)$$

The suitable modification is an inversion of the X axis orientation ($\pi$ phase shift) every second segment. Fig 7 illustrates the described case in the Bloch sphere representation.

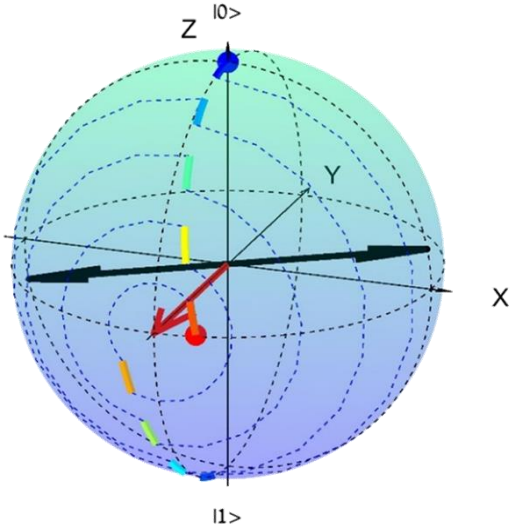

*Fig 7: Spin driving modification in the presence of interleaved dynamical decoupling with orthogonal axis. There are two types of dynamics within the Bloch spheres: Rabi driving segments are represented by colored arcs and DD driving is depicted by thin dashed arcs and red axis. The rotation axes are black for the Rabi driving segments and red for the DD.*

### IV. Alternating $B_Z$ magnetic field in the SNRG protocol

As noted above, in the SNRG scheme, we alternate the magnetic field generated detuning $\Delta_Z$ between positive and negative values after each DD $\pi$ pulse. We do so in order to constructively accumulate its effect from multiple driving segments. During the DD pulses themselves, we zero the magnetic field to drive the on - and off-resonance spins. The exact time dependent form of the magnetic field is a rectangular pulse train function (Fig 8):

$$U(t) = \begin{cases} +1, & T_{4i} < t < T_{4i+1} \\ 0, & T_{4i+1} < t < T_{4i+2} \\ -1, & T_{4i+2} < t < T_{4i+3} \\ 0, & T_{4i+3} < t < T_{4i+4} \end{cases} \quad (B7)$$

The switching timings of the magnetic field are:

$$T_i = \begin{cases} 0, & i = 0 \\ \bar{\tau} \cdot \dfrac{i}{2} + \epsilon \cdot \dfrac{i-1}{2}, & 1 \leq i \leq 8N-1, i\ odd \\ \bar{\tau} \cdot \dfrac{i}{2} + \epsilon \cdot \dfrac{i}{2}, & 1 \leq i \leq 8N-1, i\ even \\ \bar{\tau}(8N), & i = 8N+1 \end{cases} \quad (B8)$$

Where $\bar{\tau} = \dfrac{\tau}{8N}$ is the duration of a single driving segment, and $\epsilon$ is the DD pulse duration. We choose a symmetric version of XY-8 sequence for higher order noise canceling.

The corresponding microwave frequency is switched between the values $f = \gamma(B_0 \pm B_1)$ every segment, and the DD frequency is constantly set on $f = \gamma B_0$. Thus, all the pulses are aligned exactly to X and Y axes in the changing rotating frame.

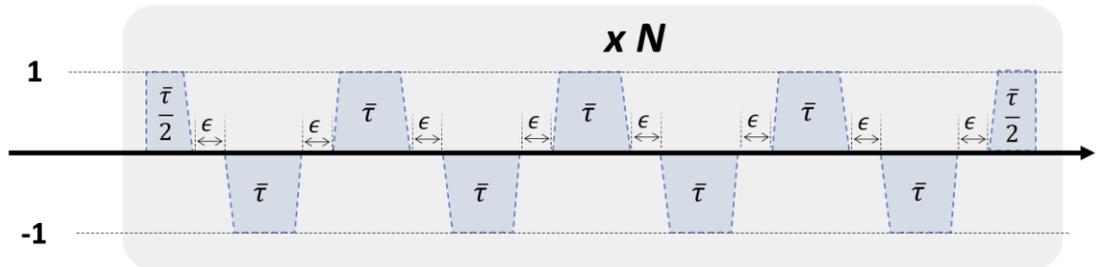

*Fig 8: $U(t)$ as a rectangular pulse train function.*

## V. Rabi driving under changing magnetic field

As mentioned in the main text, in the SNRG scheme, the detuning $\Delta_Z$ is alternating between positive and negative values after each DD $\pi$ pulse. This is done in order to constructively accumulate the phase from the multiple driving segments.

The phase changes induced by the varying magnetic field $B_Z$, require modification to the Rabi driving scheme.

We developed a modified model for the qubit's control under these conditions. The model is described by a time-dependent Hamiltonian $H_0 = \hbar\gamma B_Z(t)S_Z$ .where $\gamma$ is the spin gyromagnetic factor. At different times, the Hamiltonian commutes with itself. Hence, the time-evolution operator can be expressed as

$$U(t) = e^{-i\gamma \int_0^t B_Z(s)S_Z ds} = e^{-i\gamma \left(\int_0^t B_Z(s)ds\right)S_Z} \quad (B9)$$

Therefore, the suitable driving term (around a perpendicular axis X) is given by $H_1 = \hbar\Omega \cos\left(2\pi\gamma \int_0^t B_Z(s)ds\right)S_X$ where $\Omega$ is the field amplitude (in angular frequency units). Equivalently, the driving $B_X$ field is described by the frequency modulation equation:

$$B_X(t) = \gamma^{-1}\Omega \cos\left(2\pi\gamma \int_0^t B_Z(s)ds\right) \quad (B10)$$

The concept is elucidated with a basic example of a piecewise constant magnetic field:

$$B_Z = \begin{cases} B_1, & 0 < t < T \\ B_2, & t > T \end{cases} \quad (B11)$$

The instantaneous phase for $t > T$ is

$$\gamma \int_0^t B_Z(s)ds = \gamma B_1 T + \gamma B_2 \cdot (t - T) = \gamma B_2 t + \gamma(B_1 - B_2) \cdot T \quad (B12)$$

Hence, the respective $B_x$ driving field is:
$$\gamma B_X = \begin{cases} \Omega \cos(2\pi \cdot ft), & 0 < t < T \\ \Omega \cos[2\pi \cdot (f + \delta f)t - \delta\phi], & t > T \end{cases},$$
with $f = \gamma B_1, \delta f = \gamma(B_2 - B_1), \delta\phi = 2\pi\delta f \cdot T$. (B13)

Thus, in order to maintain uninterrupted rotation despite the change in the $B_Z$ field one needs to vary both the frequency and the phase of the driving field, but still retain the same field amplitude.

.

## Appendix C: Additional experimental details

### I. Calibration of microwire generated magnetic field

As part of DPG and SNRG schemes, we used a magnetic field gradient, generated by a current flowing through the nearby microwire. The magnetic field strength is calibrated by measuring the resonance frequency shift in an optically detected magnetic resonance (ODMR) experiment, as a function of the current in the microwire (see Fig 9). The current was fed straight from the AWG device, without any amplification. The magnetic gradient was deduced from the measured distance of the center from the microwire. Its strength is much lower than the estimated requirement for distinguishing coupled NV's. However, it can be straightforwardly enhanced by reducing the wire's diameter, and by using current amplification.

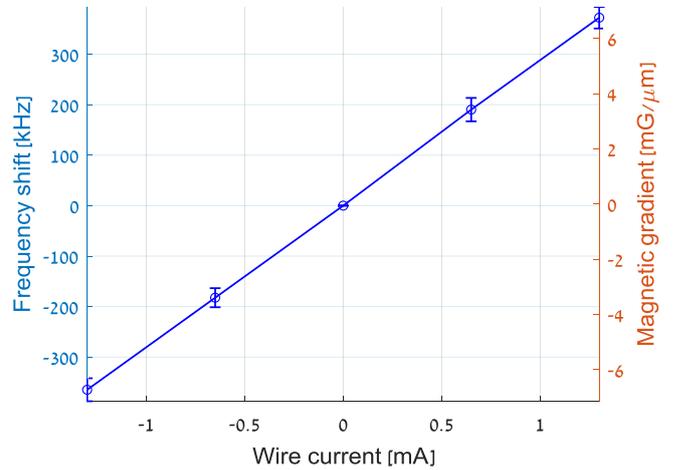

*Fig 9: Resonance frequency shift and magnetic gradient strength as a function of the current in the microwire. The NV center is located 20 μm from the wire's center.*

## II. Noise Calibration

In our experiments, the qubit decoherence is dominated by slowly fluctuating magnetic noise (as described in the main text), which can be analytically approximated by the Ornstein - Uhlenbeck (OU) process. We exploit this model to study the performance of our schemes. To do so, we search for the appropriate experimental parameters, bath coupling strength - $b$ and bath correlation time - $\tau_C$, by scanning through them. The fitted parameters are $b = 42 kHz$, $\tau_C = 230\ \mu s$. A simple continuous Rabi oscillation with the fitted parameters is shown in Fig 10. It is in fact a horizontal slice from the left panel of Fig 4.

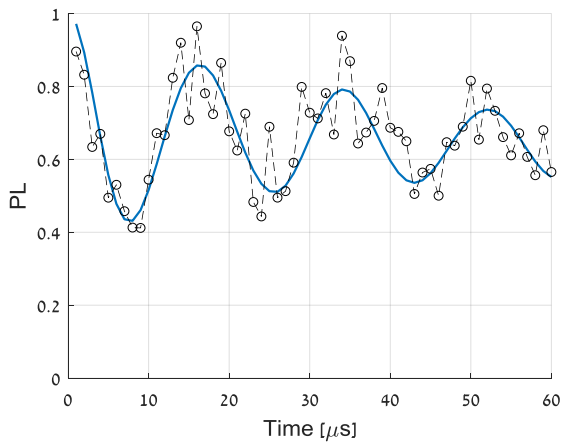

*Fig 10: Ornstein - Uhlenbeck noise parameters fitting. Comparison between NV measurement (black dots) and numerical simulation (blue line).*

Besides, we model the quality of our DD π pulses, for better simulating complex SNRG scheme. For that, we use one fitting parameter, pulse imperfection $\sigma_{DD}$, which indicates a random multiplicative error of the rotation angle. After scanning the SNRG scheme with various values, we fitted $\sigma_{DD} = 0.085$. The fit is presented in Fig 11 (it is a horizontal slice from the right panel of Fig 4).

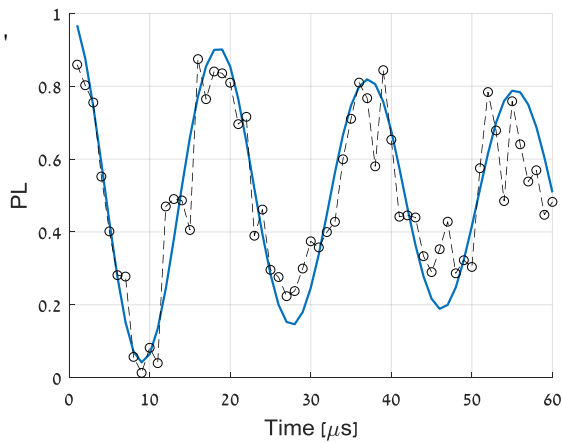

*Fig 11: DD imperfection parameter fitting. Comparison between NV measurement (black dots) and numerical simulation (blue line).*

## III. Simulative extrapolation for extreme Rabi frequencies

Utilizing numerical modeling, it is possible to foresee the performance of the various schemes for lower Rabi frequencies. Here, in Fig 12, we show a two-dimensional scan (detuning strength and Rabi frequency) of the output state fidelity vs the detuning and the driving strength. From these calculations, we extract the on resonance fidelity and bandwidth measures which are presented in Fig 4.

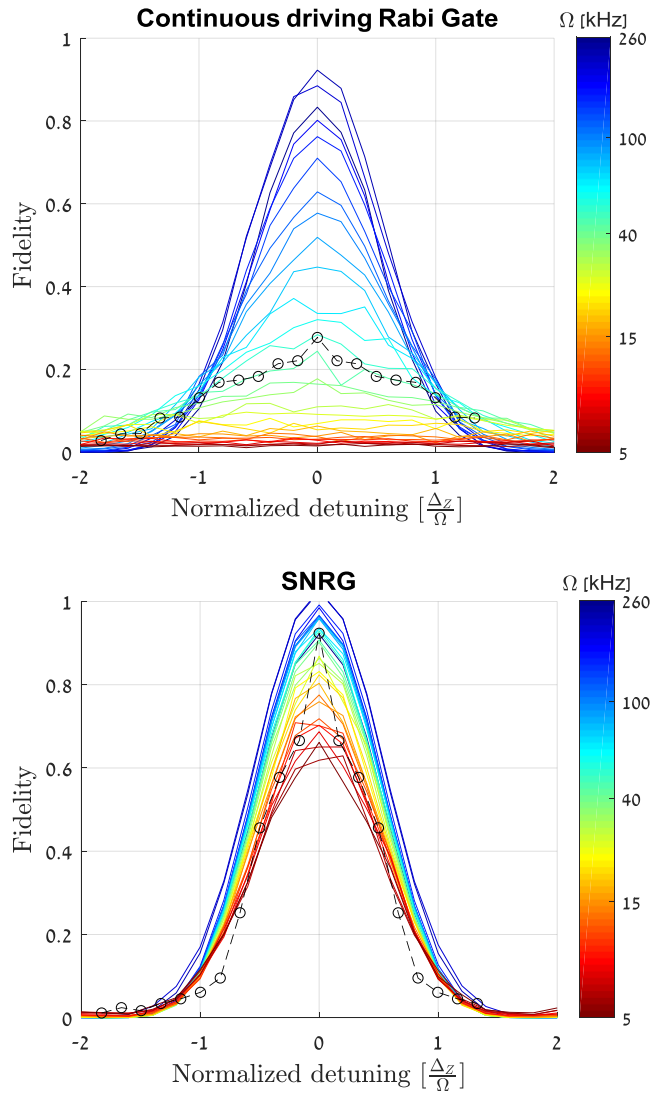

*Fig 12: Scanning of the detuning and Rabi frequency for both schemes: simple Rabi gate (top) and SNRG scheme (bottom). The dotted curve represents the measured data with $\Omega = 54\ kHz$.*

All relevant data is available from the authors upon request.